\documentclass{elsart}
\usepackage{epsfig}
\begin{document}

\begin{frontmatter}
\title{ A Search for Fine Structure of the Knee in EAS Size Spectra}

\author[privat]{G. Schatz\thanksref{email}}\relax
\thanks[email]{e-mail: bgschatz@t-online.de}
\address[privat]{Habichtweg 4, D-76646 Bruchsal, Germany}

\maketitle

\begin{abstract}
28 size spectra of extensive air showers in the knee region 
from 7 different experiment are analysed consistently. 
They are fitted by adjusting either 4 or 5 parameters: 
knee position, power law exponents above and below the knee 
region, overall intensity and, in addition, a parameter 
describing the smoothness of the bend. The residuals are then 
normalized to the same knee position and averaged. 
When 5 parameters are employed no systematic deviation 
from a simple smooth knee is apparent at the 1 \% level 
up to about a factor of 4 above the knee. At larger 
shower sizes a moderately significant deviation can 
be seen whose shape and position are compatible with 
a second knee caused by iron group nuclei.
\end{abstract}
\end{frontmatter}

\section{Introduction}
The existence of the 'knee' in the spectra of extensive air showers (EASs) has 
been known by now for more than 40 years \cite{kul}. First seen in 
the number of electrons (the shower 'size') observed near sea level 
it was later also observed in the muon number \cite{stam,durban}, hadron properties
\cite{yosh,hoer} and muon densities \cite{hau}. 
In fact it seems to show up in all shower observables 
if investigated in sufficient detail. Nevertheless its origin is still 
obscure. Of the explanations proposed two seem to have found more 
general acceptance.\\

\begin{figure}
\begin{center}
\epsfig{file=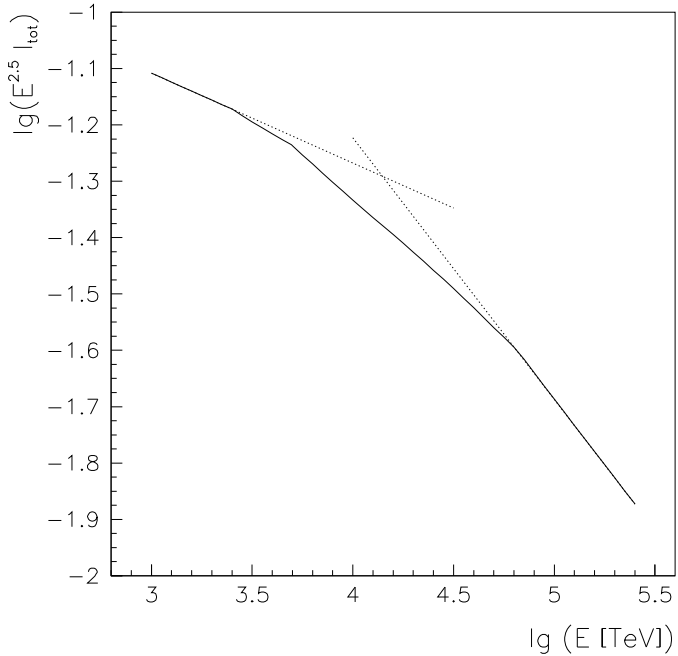,width=10cm}
\caption{Theoretical all particle energy spectrum of cosmic 
rays obtained by extrapolation from lower energies and 
assuming knee energies increasing in proportion to the 
nuclear charge. The vertical scale is in arbitrary units.}
\label{wiebsk}
\end{center}
\end{figure}

The first of these relates the knee to the influence of interstellar 
magnetic fields during propagation of the cosmic ray particles 
in or leakage from the Galaxy or in the course 
of acceleration for which magnetic fields 
probably play a major role. Since the radius of curvature of an extremely 
relativistic particle in a magnetic field is proportionate to its 
energy $E$ and inversely proportionate to its nuclear charge $Z$ one 
would expect the knee then to show up at different energies for 
different nuclear species among the primary particles. In fact one 
would expect the energy spectra of each element to show a 
knee at an energy displaced 
by a factor of $Z$ with respect to that of protons. This was 
probably first realized by Peters \cite{pet} who discussed 
the implications on various shower observables in great 
detail\footnote{I thank P. Grieder (Bern) for drawing my 
attention to this important old reference.}. The effect of this 
on the total energy distribution of cosmic ray particles is illustrated 
in fig. \ref{wiebsk} which shows the result of a simple model calculation 
based on data compiled by Wiebel-Sooth et al. \cite{wiebel}. 
These authors give, for each element up to nickel ($Z=28$), 
the differential flux at 1 TeV and the 
exponent of the power law spectrum. 
These partial spectra were extrapolated up to a knee which was assumed 
to lie at $lg (E_{K,Z}[TeV]) = 3.4 + lg Z$. This corresponds to a knee
energy of c. 2.5 PeV for protons. It was then assumed that the 
exponent increased abruptly at the knee by 0.35 for all elements. All 
statistical errors (which are of course substantial owing to the 
extrapolation by some orders of magnitude) were neglected. 
The full line in fig. \ref{wiebsk} 
represents the sum of all contributions 
from hydrogen to nickel. The two dotted lines are the 
extrapolations of the 
all particle spectrum from below the proton and above the iron 
knee. As fig. \ref{wiebsk}  
shows the change of slope is, unsurprisingly, not confined to 
one position or small range along the line. At least three bends 
are obvious corresponding to hydrogen, helium and the  
iron group. Hence in this case one would clearly expect 
a more complicated structure of the spectrum than a simple bend 
for all shower observables depending on primary energy.\\

The second proposal, less popular than the first one though 
recently reasserted by Nikolsky and Romachin \cite{nikrom}, 
attributes the origin of the knee to the properties of high energy 
interactions in the atmosphere. A change of the spectrum of 
observables on ground level might of course occur if strong 
interaction changes by some kind of threshold phenomenon. 
The knee energy is estimated to be near a few PeV which is 
approximately a factor of 2 above the highest centre of mass 
energy available in the laboratory today. Therefore such an 
effect cannot at present be excluded.  
Since an EAS induced by a nucleus of mass 
number $A$ and energy $E$ may, to a reasonable 
first approximation, be considered as a superposition of 
$A$ showers induced by nucleons of energy 
$E/A$ (superposition principle) 
one would expect, under this assumption, a similar shift of the 
knee position as described above but by a factor of $A$ instead of $Z$. 
Again a more complex structure of the knee would 
appear natural and a picture very similar to the one shown 
in fig. \ref{wiebsk} would be obtained.\\

It should be mentioned that the KASCADE collaboration has 
recently presented evidence that the knee is to be attributed 
to light nuclei and that the spectrum of heavier nuclei 
does not exhibit a change of slope 
in the vicinity of  the 'main' knee. This 
claim is based on a comparison of electron and muon numbers 
of EASs with Monte Carlo simulations \cite{slc} but also on 
a phenomenological classification of EAS by their electron 
to muon ratios \cite{hau}.\\

Erlykin and Wolfendale have recently, in a series of papers \cite{ew}, 
claimed observational evidence for a more complicated structure 
of the knee region. This structure which according to the authors 
does not show up clearly in single measurements but becomes 
visible when averaging several ones, is attributed to the 
influence of a recent nearby supernova. It is probably not 
unfair to say that not many have been convinced by the 
empirical evidence claimed by the 
authors. But the underlying idea 
appears very intriguing and reasonable. 
The solar system has, during the 4.5 
billion years of its existence, probably been passed by 
several if not many shock fronts from supernovae 
exploding in its vicinity. One of these clearly must be 
the most recent one and it appears well conceivable that 
the cosmic ray spectrum which we observe today is influenced 
by this individual event (and hence not typical for 
the whole Galaxy). The possibility of a single source 
having a large impact on the energy and mass 
distributions of cosmic rays at the earth was already 
realized by Peters \cite{pet}.\\

For all these reasons it appears interesting to 
study the shape of the knee region in more detail 
and to look as to whether it can be really described by 
a simple bend between two power laws 
or whether any of the effects mentioned above 
can be identified. Except for the work by Erlykin 
and Wolfendale the author is not aware of any other attempt 
in this direction. A definite negative result would 
not only worry some authors but also present 
difficulties for the usual models of the origin 
of the knee whose existence, on the other hand, 
is beyond any doubt.\\

In this paper I attempt to compare 28 different 
measured spectra of the electron number $N$ in 
the knee region. (I drop the usual 
subscript $e$ because there can be no confusion in this paper.) 
The data originate from 7 
experiments and cover a range of atmospheric depths 
between 730 and 1250 [$g/cm^{2}$]. The electron 
number (or shower 'size') is probably the shower 
observable for which the largest amount of 
measurements exist. The basic procedure adopted is 
the following: Each spectrum is first fitted 
separately by an adequate function adjusting either 
4 or 5 parameters. In a next step 
the residual spectra are shifted to the same 
knee position and averaged. It may be expected 
that this averaging reduces not only the 
statistical fluctuations of the measurements but 
also (at least part of) the systematic ones and hence should 
make any deviations from the pre-chosen fit  function 
more conspicuous. \\

The data used and their analysis are described 
in the following section. In section 3 we compare 
the results of the analysis with a simple model 
and present our conclusions in section 4.\\

\section{Data and analysis}
\subsection{The data base}
A list of the spectra analysed in this paper and 
their sources can be found in table 1. The EAS-TOP 
experiment is the only one which has published \cite{eastop} the 
$N$ spectra in numerical form\footnote{It should be mentioned 
that the scale factor quoted in the caption of the relevant 
table 1 of ref. \cite{eastop} should read $10^{-7}$ 
instead of $10^{-8}$ \cite{gianni}.}. The CASA data were 
read from table 6.2 of reference \cite{glas} (which is 
equivalent to fig.  of ref. \cite{gla}). Such a 
procedure is of course of limited accuracy and does not exhaust 
the statistical precision of the data (especially at low 
shower sizes). All other data sets were made available 
in numerical form by the authors\footnote{I thank A. 
Chilingarian (Yerevan), R. Glasstetter 
(Karlsruhe), G. Heinzelmann (Hamburg), N. N. Kalmykov (Moscow) 
and M. Nagano (Tokyo) for their invaluable support.}. Table 1 
gives some details of the data together with the quality 
of the two kinds of fits described in the next subsections. 
Several experiments registered events in different 
ranges of zenith distance which then of course 
correspond to different atmospheric depths. 
The total data set comprised 784 points which is to be 
compared with a total of 112 or 140 derived parameters 
for the four or five parameter fits, respectively.

\begin{table}
\center
{\small
\begin{tabular}{|c|c|c|r|r|r|r|c|}
\hline
  & atmospheric & number &\multicolumn{4}{|c|}{quality of fit} &  \\
Experiment & depth & of data &\multicolumn{2}{|c|}{4 parameters} &\multicolumn{2}{|c|}{5 parameters} & ref. \\
   & [$g/cm^{2}$] & points & \multicolumn{1}{|c|}{$\chi^{2}$} & $\chi^{2}/F$ & \multicolumn{1}{|c|}{$\chi^{2}$} & $\chi^{2}/F$ &  \\ 
\hline
     & 928 & 41 & 52.41 & 1.416 & 52.41 & 1.456 & \\
     & 1021 & 40 & 112.87 & 3.135 & 112.87 & 3.225 & \\
\raisebox{1.5ex}[-1.5ex]{AKENO} & 1114 & 36 & 39.33 & 1.229 & 39.33 & 1.269 &
\raisebox{1.5ex}[-1.5ex]{\cite{nag,nag2}}\\
     & 1206 & 34 & 51.41 & 1.714 & 51.41 & 1.773 & \\
\hline
     & 883 & 20 & 37.80 & 2.362 & 15.53 & 1.017 &  \\
     & 911 & 20 & 35.63 & 2.227 & 19.41 & 1.137 &  \\
     & 941 & 20 & 58.02 & 3.627 & 43.29 & 1.699 &  \\
CASA & 972 & 20 & 34.63 & 2.164 & 29.40 & 1.400 & \cite{glas} \\
     & 1006 & 20 & 29.65 & 1.853 & 18.87 & 1.122 &  \\
     & 1042 & 20 & 47.84 & 2.990 & 32.73 & 1.477 &  \\
     & 1081 & 20 & 29.56 & 1.847 & 28.85 & 1.387 &  \\
\hline
     & 835 & 30 & 26.51 & 1.020 & 26.51 & 1.061 & \\
     & 880 & 30 & 17.82 & 0.685 & 13.98 & 0.559 & \\
     & 920 & 30 & 13.42 & 0.516 & 13.12 & 0.525 & \\
\raisebox{1.5ex}[-1.5ex]{EAS-TOP} & 960 & 30 & 20.34 & 0.782 & 16.77 & 0.671 & \raisebox{1.5ex}[-1.5ex]
{\cite{eastop}}\\
     & 1000 & 30 & 6.14 & 0.236 & 6.14 & 0.246 & \\
     & 1040 & 30 & 9.91 & 0.381 & 7.96 & 0.318 & \\
\hline
HEGRA & 820 & 12 & 16.76 & 2.095 & 16.68 & 2.383 & \cite{hein}\\
\hline
        & 1047 & 33 & 87.62 & 3.021 & 42.80 & 1.529 & \\
        & 1098 & 34 & 42.39 & 1.413 & 26.51 & 0.914 & \\
KASCADE & 1149 & 34 & 39.11 & 1.304 & 25.57 & 0.882 & \cite{durban,rgl}\\
        & 1200 & 33 & 54.80 & 1.890 & 24.97 & 0.892 & \\
        & 1251 & 30 & 31.64 & 1.217 & 25.53 & 1.021 & \\
\hline
          & 731 & 30 & 58.51 & 2.250 & 52.84 & 2.114 & \\
          & 804 & 28 & 19.74 & 0.822 & 19.69 & 0.856 & \\
\raisebox{1.5ex}[-1.5ex]{MAKET-ANI} & 876 & 27 & 27.00 & 1.174 & 26.93 & 1.224 & \raisebox{1.5ex}[-1.5ex]
{\cite{ashot}}\\
          & 949 & 27 & 31.28 & 1.360 & 28.41 & 1.292 & \\
\hline
MSU    & 1068 & 25 & 85.31 & 4.062 & 84.84 & 4.142 & \cite{nick}\\
\hline
\end{tabular}}
\normalsize
\caption{Data sets used in this analysis and fit results. 
The number of degrees of freedom F is the number of 
data points minus 4 or 5, respectively.}      
\label{data}
\end{table}

\subsection{Four parameter fits}
All fits performed were least squares fits with $lg N$ and 
$lg (N^{2.5}I)$ as the variables and employing the usual 
weights derived from the quoted errors. Here $I$ is the 
differential flux. In  a first step, 
the fit function chosen consisted of two power laws connecting 
continuously at the knee position:

\begin{equation}
I(N) = I_{K} \left( \frac{N_{K}}{N}\right) ^{\gamma}
\end{equation}

Here $I_{K}$ is the differential flux at the knee position and
$\gamma = \gamma_{1}$  for  $N < N_{K}$  and  $\gamma = \gamma_{2}$  for  $N > N_{K}$.

\begin{figure}
\center
\epsfig{file=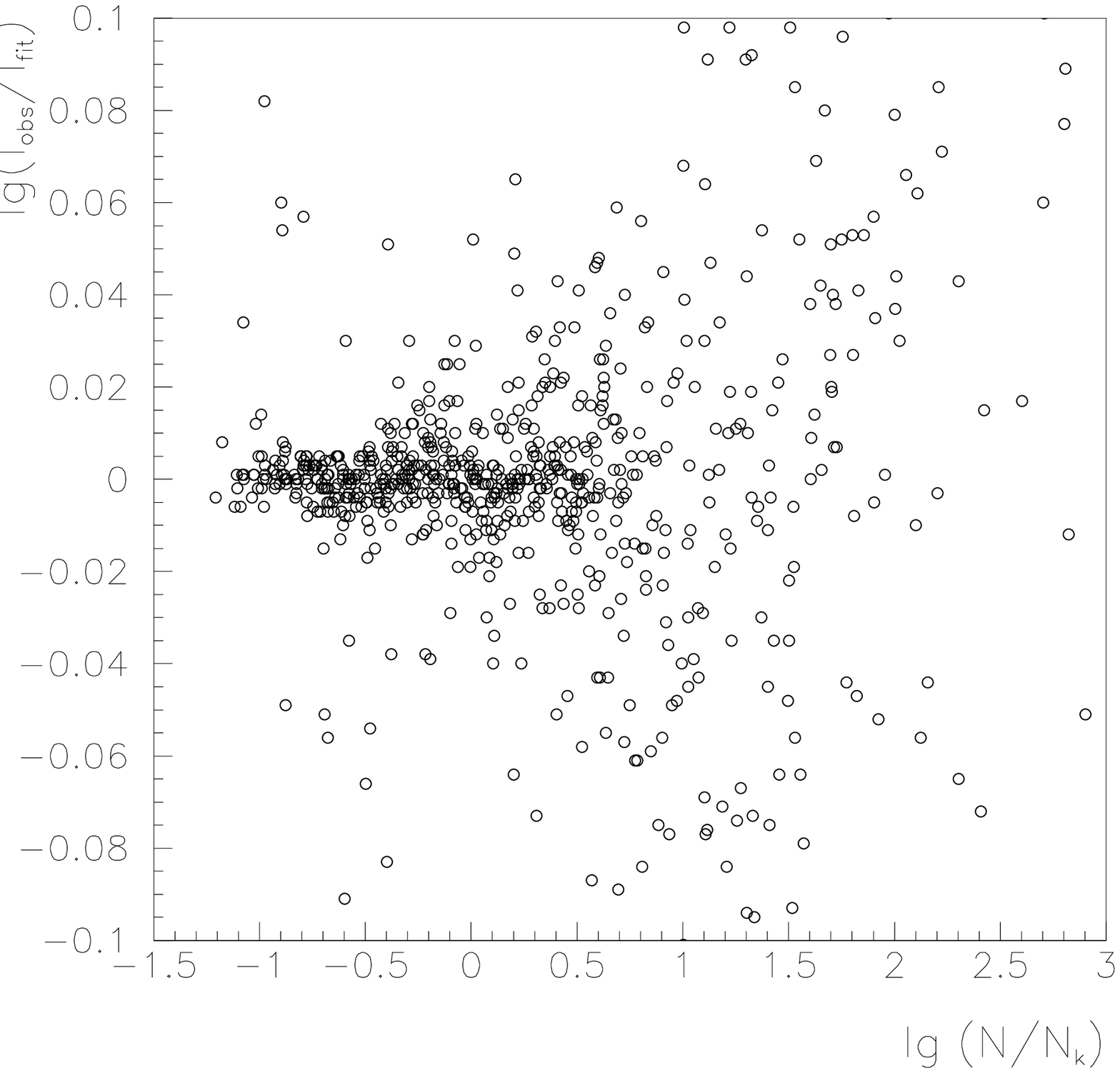,width=10cm}
\caption{Differences between observed and fit values of 
the differential flux normalized to the same knee 
position. The data are based on fits with four parameters. 
Each symbol represents one data point from 
one of the 28 spectra. No error bars have been drawn in 
order not to confuse the picture but most are 
compatible with 0.}
\label{diffig}
\end{figure}

The quantities $I_{K}, N_{K}, \gamma_{1}$ and $\gamma_{2}$
were adjusted to describe the data. 
After the fit, the differences between 
observed and fit values were calculated for each of the 
28 spectra. These spectra of residuals were then shifted 
along the $lg N$ axis to the same knee position. 
This amounts to choosing $lg (N/N_{K})$ as the new independent 
variable where $N_{K}$ is the knee position found for the 
respective spectrum. 
The scatter of the data points after these procedures is 
illustrated in fig. \ref{diffig}  which clearly exhibits the statistical 
nature of the residuals (and the increase of the 
errors with increasing $N$). No errors bars have been drawn because this 
would completely confuse the picture but most of them 
are compatible with 0. \\

As the next step, 
all data points within horizontal intervals of width 0.1 
were averaged neglecting the horizontal uncertainties 
resulting from the errors of $N_{K}$. The latter were smaller 
than the chosen bin width for almost all spectra. 
The number of data points within a given interval was above 30 
in the vicinity of the knee and dropped to near 1 at 
the extreme ends of the total range because the $N$ ranges 
of the various measurements did not coincide. 
It should be mentioned that most of the original data 
were also binned in intervals of 0.1 width so this choice 
was very natural. \\

\begin{figure}
\center
\epsfig{file=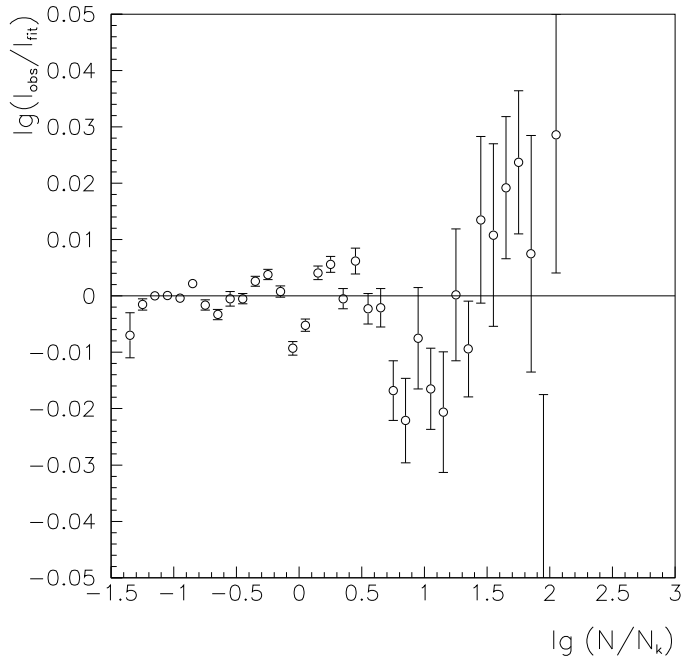,width=10cm}
\caption{Averaged differences between observed and fit values 
as shown in fig. \ref{diffig} for the fits with 4 parameters.}
\label{dif4pmit}
\end{figure}

The averaging procedure described actually amounts to 
taking the geometric means 
of $I_{obs}/I_{fit}$ (with adequate weights). This 
eliminates all sensitivity to the absolute normalization 
of the data. There is 
reason to suspect that the quantity which different 
experiments call shower size $N$ is not the same. 
This is most obvious when the influence of muons 
on the experimental results is considered. The 
electron detectors, in all experiments scintillation 
counters, are also sensitive to muons. Some 
experiments, such as KASCADE \cite{kas,web}, put a lot of 
effort into correcting for this effect. This is 
impossible, on the other hand, if 
muons are not measured independently as is the case in 
some of the experiments included in this analysis. 
Also different detector thresholds or saturation 
effects may lead to systematic differences. The 
advantage of choosing $lg (N/N_{K})$ as a new variable 
is that constant scale factors cancel. 
Although not all of the effects mentioned above 
will lead to modifications of $N$ by a constant 
factor at least some instrumental effects will 
be reduced or even removed. \\

The weighted averages thus obtained are displayed in fig. 
\ref{dif4pmit} with their 
statistical errors (which have been multiplied, as usual, 
by the root of $\chi^{2}/F$ if the latter was larger 
than 1). A number of points at the upper end of the spectrum 
have been omitted here because of their huge errors. These  
were all compatible with 0 within their statistical errors. In spite of the 
considerable errors the mean residuals can hardly be said 
to scatter statistically. The most 
significant deviations, at the 
two points neighbouring the knee position, amount to 
$7.6\sigma$ and $4.6\sigma$ and are 
therefore highly significant. \\

A large discrepancy in the immediate vicinity of the knee 
position is in no way surprising because several 
experiments have shown that the slope of the spectrum 
does not change abruptly at the knee but rather 
smoothly over a finite range of $lg N$. The sign of the 
deviation is precisely as would be expected in such 
a situation. So the most prominent deviation just 
reflects a well known and expected feature. On the other hand, 
this result also shows that the procedure adopted 
in this paper is suitable to reveal more clearly a 
structure where it is present. But obviously a more suitable 
fit function has to be employed.\\

It is worthwhile mentioning that the knee positions 
resulting from the fits show the well known dependence 
on atmospheric depth. Five of the seven experiments 
with 18 spectra are in good agreement with a simple linear 
dependence. The other two can be made to agree 
when the shower sizes are scaled down by 30\% and 70\%. 
The latter would not affect the present analysis in any way, of course. 
The scatter of the other parameters is clearly larger than the 
quoted statistical errors would indicate. This is subject 
of further analysis. Since in this study we are focussing on 
deviations from a smooth size dependence in a limited size range 
I would not expect such (apparently systematic) differences 
between experiments to influence my conclusions.\\

\subsection{Five parameter fits}
Several different methods have been used to describe 
a smooth knee all of which seem to work reasonably 
well. A very popular procedure is to select 
a finite interval in the vicinity of the knee and 
then first fit power laws to the regions above and 
below this interval. In a second step these functions 
are then connected smoothly, e.g. by a polynomial. 
A disadvantage of this method is, in my opinion, 
that at least two but usually more additional 
parameters are introduced (which does not seem to be 
always realized). \\

Ter-Antonyan and Haroyan \cite{ter} have used a 
function which avoids this disadvantage. 
Their expression

\begin{displaymath}
I(N)=I_K\cdot \left(\frac{N}{N_{\rm K}}\right)^{-\gamma_1} \cdot
        \left(1+\left(\frac{N}{N_{\rm K}}\right)^\varepsilon\right)
                ^{(\gamma_1-\gamma_2)/\varepsilon}        
\end{displaymath}

can be shown to approach pure power laws far from 
the knee and to tend to eq. (1) for $\epsilon \rightarrow 0$. 
The additional parameter $\epsilon$ does not lend itself 
to a direct physical interpretation, though. Also the 
two regions above and below the knee are not 
described in a formally equivalent way.\\

I have adopted a different approach. It consists 
of folding eq. (1) with a normalized distribution of 
finite variance. This implies a model in which a
hypothetical underlying spectrum with a sharp knee 
is observed with finite resolution. The exact shape 
of the 'resolution function' is not really important 
but if a log-normal distribution is chosen the following 
closed expression is obtained:

\begin{equation}
I(N) = I_{K} \left( e^{\sigma^{2}\gamma^{2}_{1}/2} ~\Phi(u_{1}) \left( \frac{N_{K}}{N}\right) ^{\gamma_{1}} 
+ ~e^{\sigma^{2}\gamma^{2}_{2}/2} ~[1 - \Phi(u_{2})] \left( \frac{N_{K}}{N}\right) ^{\gamma_{2}}\right)
\end{equation}

\begin{displaymath}
u_{i} = \sigma \gamma_{i} - \frac{ln (N/N_{K})}{\sigma}
\end{displaymath}

Here $\sigma$ is the standard deviation of the 
log-normal distribution and $\Phi(u)$ is the error integral:

\begin{displaymath}
\Phi(u) = \frac{1}{\sqrt{2 \pi}} \int \limits_{-\infty}^{u} e^{-x^{2}/2} dx
\end{displaymath}

It is again straightforward to show that eq. (2) 
approaches power laws at sufficient distances from 
the knee and tends to eq. (1) for $\sigma \rightarrow 0$. 
As far as the physical interpretation is concerned one 
should realize that $\sigma$ does not primarily represent 
instrumental effects. It is well known from 
simulations of EAS development in the atmosphere 
that the shower size $N$ fluctuates considerably 
for events of fixed primary energy, mass and direction 
of incidence. The effects of these 
fluctuations are also incorporated in the parameter 
$\sigma$. \\

I should like to point out that it is, in my opinion,  very much a 
matter of taste which of the described approaches is 
chosen. Since all describe the experimental spectra 
well within observational uncertainties 
they will all lead 
to the same conclusions. \\

\begin{figure}
\center
\epsfig{file=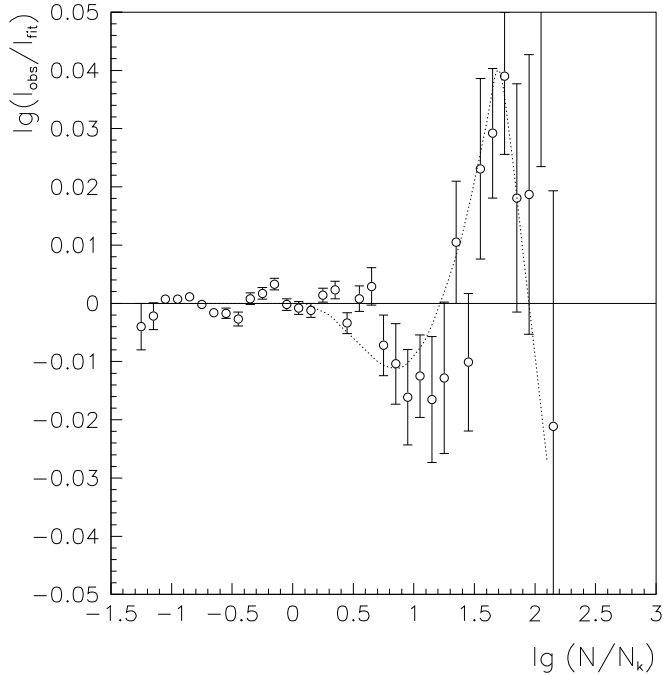,width=10cm}
\caption{Averaged differences between observed and fit values 
based on fits with 5 parameters. The dotted line is 
discussed in section 3.}
\label{dif5pmit}
\end{figure}

The fitting and averaging was then performed as 
described in the preceding subsection. Fig \ref{dif5pmit} displays 
the result. The residuals are now below 0.005 (which 
corresponds to a difference between fit and 
measurement of less than $\simeq1 \%$) for 
the whole lower part of the spectrum, up to 
$lg (N/N_{K})\simeq 0.6$. \\

The remaining deviations in the upper part 
amount to $\simeq10 \%$ on the positive 
side and $\simeq5 \%$ on the negative side.  
Their regular pattern with increasing size can 
hardly be called statistical 
in spite of the fact that only two of the last 15 
data points differ from 0 by more than 2 standard 
deviations. A possible explanation is discussed 
in the next section where the origin of the dotted 
curve in fig. \ref{dif5pmit} will also be explained. \\

One might argue that several data 
points in the lower size range clearly differ from 0 
by a similar amount statistically 
or even recognise an oscillatory structure. 
But this deviation is 
on the 1 \% level and hence a factor of 5 to 10 
smaller than the one in the upper size range. I would 
tend to argue, to the contrary, that an agreement 
between various experiments on this level is 
rather impressive and would hesitate to draw 
far reaching conclusions from these observations.\\

\section{Discussion}
The results displayed in fig. \ref{dif5pmit} show 
clearly that the lower part of the observed size 
spectrum is well describeded by two power 
laws and a simple knee. I find it 
difficult to believe, on the other hand, that 
the discrepancy visible 
in the upper part of the size range is purely 
statistical, in spite of the large errors of 
the individual points (but this is of course left 
to the reader's own judgement). It is much more 
difficult to assess possible systematic errors. 
I will come back to the latter question later in this section 
and first turn to a possible explanation of the observed 
deviations assuming they are real.\\

\begin{figure}
\center
\epsfig{file=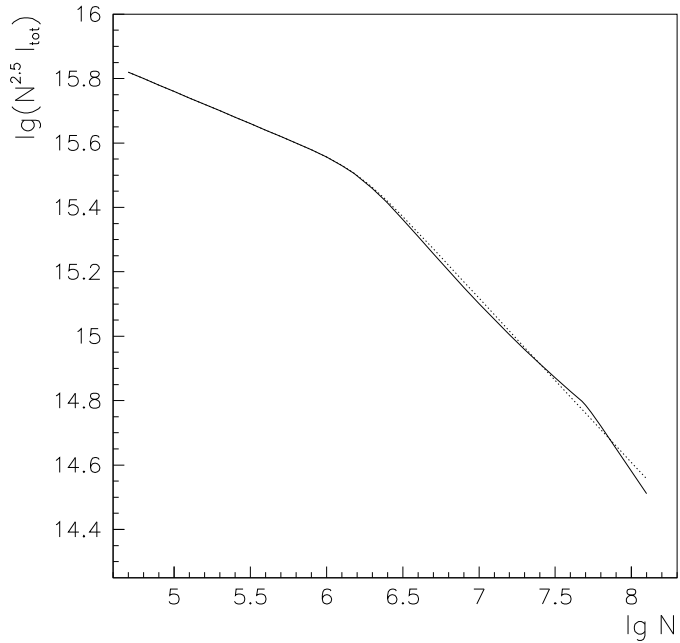,width=10cm}
\caption{Calculated size spectrum consisting of two 
functions of the type shown by eq. (2) but displaced 
horizontally (full line). The dotted curve is of the 
same type and its parameteres have been adjusted 
to give an approximate description of the full line. 
The vertical scale is in arbitrary units.}
\label{schem}
\end{figure}

Let us suppose that the real size spectrum exhibits 
two knees displaced by one or two orders of magnitude 
with respect to each other. This is not unreasonable 
in view of the considerations extended 
in the introduction and illustrated in fig. \ref{wiebsk}. 
Then a situation might occur which is sketched in fig. 
\ref{schem}. Here the full line represents a hypothetical 
spectrum with two knees. If one tries to fit this 
spectrum with an expression allowing only one knee 
this will be adjusted to the lower one because 
it is more pronounced. Also statistical errors of the data 
points increase with increasing shower size and hence 
the lower points carry larger  
weights. But the deviation of the 
data from a pure power law in the higher size 
range will influence the fit and might lead to a 
situation illustrated by the dotted line in fig. 
\ref{schem}. This would then result in differences between 
data and fit which agree qualitatively with those 
observed in fig. \ref{dif5pmit}. I admit that the 
separation of the full and dotted lines in fig. 
\ref{schem} does not look dramatic but one should 
be aware of the difference in the vertical 
scales between figs. \ref{dif5pmit} and \ref{schem}.\\

It is possible to extend this conjecture a step 
further. For this let us try to estimate the {\it size}  
spectrum resulting from an {\it energy} 
spectrum as displayed in fig. \ref{wiebsk}. Essentially 
all simulations of EASs lead to a power law relation 
between shower size and energy for primary protons, 
at least in the range of energy and atmospheric 
depth relevant for this analysis:

\begin{displaymath}
N_{p} = a ~E^{\delta}_{p}
\end{displaymath}

The corresponding relation for a heavier primary 
may then be estimated by help of the superposition 
principle invoked already in the introduction:

\begin{displaymath}
N_{A} = A ~a \left( \frac{E_{A}}{A} \right) ^{\delta}
\end{displaymath}

Here $E_{A}$ denotes the energy of a nucleus of 
mass number $A$. Assuming again, as in the introduction, 
that the knee position in energy is shifted by a 
factor of $Z$, $E_{K,A}=ZE_{K,p}$, one obtains 
for the shift of the knee in shower size

\begin{displaymath}
N_{K,A} = A ~a \left( \frac{E_{K,A}}{A} \right) ^{\delta} = a ~Z^{\delta} ~A^{1-\delta} ~E^{\delta}_{K,p}
\end{displaymath}

\begin{displaymath}
lg N_{K,A} = lg N_{K,p} + \delta ~lg Z -(\delta -1) ~lg A
\end{displaymath}

So in this simple model the shift depends only 
on mass and charge of the nuclei. The second 
column of table \ref{knie} shows these shifts 
for some abundant elements. The value of 
$\delta \simeq 1.30$ which is the only additional 
information needed, was 
calculated from simulation results reported in 
tables 20 and 26 of \cite{knapp} (for the QGS-JET 
model; the other 4 models given in the same tables 
differ from this by at most 0.04). The separation 
between hydrogen and helium is reduced as compared to 
the corresponding shift in the energy spectrum due 
to the difference in mass to charge ratio. So the 
knees of these elements would merge if the smoothness 
is taken into account. Hence in this case the knee observed in 
the size spectrum has probably to be attributed to 
the combined effect of protons and helium. 
It is obvious that the shift of the 
knee for iron is of the right order of magnitude to 
explain a deviation as seen in fig. \ref{dif5pmit}.\\

\begin{table}
\center
\begin{tabular}{|c|c|c|}
\hline
  &\multicolumn{2}{|c|}{$lg (N_{K,A}/N_{K,p})$} \\
\raisebox{1.5ex}[-1.5ex]{Element} & $E_{K} \propto Z$ & $E_{K} \propto A$ \\
\hline
He & 0.21 & 0.60 \\
C  & 0.69 & 1.08 \\
O  & 0.81 & 1.20 \\
Ne & 0.91 & 1.30 \\
Mg & 0.99 & 1.38 \\
Si & 1.06 & 1.45 \\
Fe & 1.32 & 1.75 \\
\hline
\end{tabular}
\caption{Estimated shifts of the knee 
in the size spectrum with respect to the proton knee 
under the simple assumptions that the knee energy is 
either proportionate to charge or to mass of the 
primary nuclei.}
\label{knie}
\end{table}

If the other (less popular) proposal for the 
origin of the knee is adopted, $E_{K,A}=AE_{K,p}$, 
one obtains in a similar way

\begin{displaymath}
lg N_{K,A} = lg N_{K,p} + lg A
\end{displaymath}

The corresponding values have also been entered into 
table \ref{knie}. Intriguingly, the position of the 
iron knee obtained under the 
assumption of $E_{K,A} \propto A$ is in better 
quantitative agreement with the deviations 
observed in fig. \ref{dif5pmit}. On the other hand, 
the knees originating from protons and helium 
should then appear well separated of which there 
is not much evidence in the figure. \\

In order to check this possibility further I have 
calculated the following 'toy' model. I have assumed 
that the size spectrum can be described by the sum of two 
functions of the type given in eq. (2) with 
different intensities, the same 
slopes and the two 
knees separated by $\Delta lgN_{K}=1.7$. 
This model spectrum then has two free 
parameters to adjust, the ratio of  intensities and 
the change of the power law exponents. For 
the latter $\Delta \gamma=0.5$ was assumed. The 
parameters $\sigma$ were chosen to reproduce the 
variance of the simulation results in the quoted 
tables of ref. \cite{knapp} and were 0.44 for protons 
and 0.08 for iron. This choice neglects all instrumental 
effects on the size resolution and should therefore represent 
a lower limit. This calculated 
spectrum was then fitted by a single function of the same 
type with the knee at the same position as the 
lower of the other two. Intensity and slope below 
the knee were taken 
to be the same as those of the model spectrum, and 
the slope above the knee adjusted. The difference of 
the model and fit spectra are shown in  fig. 
\ref{dif5pmit} as the dotted line. It should be 
emphasized that the width of the knees were not 
varied but kept fixed at the values quoted above. 
The three new parameters (relative intensity, change 
of the exponent of the model function and slope 
above the knee of the fit function) were adjusted to some 
extent but no serious attempt was made to obtain 
a perfect fit. For this all elements expected to be 
present in primary cosmic rays should have 
to be taken into account. Also there is no compelling 
reason to believe that all partial spectra have the 
same exponents. This would then leave more
parameters to adjust than data points in fig. \ref{dif5pmit}. 
Although the dotted line does not reproduce the data 
perfectly the model gives a reasonable description 
in view of its crudeness. Hence the data can be said 
to be in agreement with a second component in 
the overall spectrum exhibiting a knee at higher 
shower size. It remains surprising, though, that 
the displacement is clearly in better agreement 
with a proportionality of the knee position to 
mass rather than to charge 
of the primary nuclei (if the second component is 
attributed to iron which appears to be the most 
reasonable choice).\\

This is probably the right moment to stop speculating 
and to turn to the thornier question of systematic 
errors. It should be mentioned that the number 
of experimental points averaged drops from 29 at 
$lgN=0.6$ to 7 at 2.1. So the region where the 
deviation is observed covers the highest data points 
of several experiments and these might be under 
suspicion to suffer from saturation effects. Saturation 
will lead to an overestimate of the differential flux 
increasing towards the end of the range of measurement. 
Therefore it is, in my opinion, not possible to exclude 
at this moment systematic effects as a possible origin of the 
observed deviations without more detailed investigations. 
These require a better knowledge of experimental 
details and have to be performed for each experiment 
separately. This is beyond the scope of the 
present paper.\\

\section{Summary}
I have presented an analysis of 28 size spectra 
from 7 different experiments. Their results can be well 
represented by two power laws with a smooth knee 
in between up to 
a shower size a factor of $\simeq4$ above the bend. At 
larger sizes a deviation occurs which appears 
statistically sound though not overwhelming. The 
maximum of this deviation roughly coincides with 
the position where one would expect a knee originating 
from primary iron nuclei if the knee in energy is 
assumed to be shifted in proportion to primary mass. 
I realize of course that, to 
the best of my knowledge, no such feature has ever 
been seen before in {\it size} spectra although a 
second knee was seen in the derived {\it energy} spectrum 
by the AKENO group \cite{nag2}. (The deviations claimed by 
Erlykin and Wolfendale \cite{ew} are closer to the 
main knee position and cannot therefore be 
identified with the ones observed here.) 
But on the other hand 
no analysis so far has taken advantage of the 
statistics of 28 spectra from 7 experiments. 
The effect shows up in a region where 
detectors might start to saturate which may 
lead to distortions of the spectrum. Hence systematic 
errors cannot, in my opinion, presently be ruled 
out as the origin of the observed fine structure. \\

I would consider it worthwhile to study the data from 
experiments with high statistics and high 
detector quality in more detail, especially 
as far as saturation effects are concerned, in order to verify 
or not the existence of this fine structure. One way of 
checking the influence of saturation is to compare 
spectra taken by the same experiment in different 
bins of zenith distance. The measured size of a 
primary of given energy and mass decreases with 
increasing zenith angle. This results in a shift of the 
knee position to lower sizes with increasing zenith 
distance. This shift has been observed by many 
experiments and is also clearly visible in the 
data of the 5 experiments which have contributed 
more than one spectrum to this investigation. Hence 
vertical showers should saturate at lower values of 
$lg(N/N_{K})$. If 
the same structure is observed at all zenith 
distances the influence of saturation can be ruled out. 
The range of shower sizes of such a study should 
of course extend up to at least two orders 
of magnitude above the 'main' knee. 
This implies a reduction of particle flux by about 
4 orders of magnitude and hence requires very good 
statistics as well as detectors of a considerable 
dynamic range.\\

It should be mentioned that the analysis described in 
section 2 of course also yields numerical values 
for the 4 or 5 parameters and their statistical 
errors and it appears interesting to compare 
the results from the various experiments and study 
their dependence on atmospheric depth. 
This will be the subject of a forthcoming paper.\\

\noindent{\large \bf{Acknowledgements}}\\
\\
Writing this paper would have been impossible without the 
generous support of quite a number of colleagues. I thank 
especially A. Chilingarian (Yerevan), R. Glasstetter 
(Karlsruhe), G. Heinzelmann (Hamburg), N. N. Kalmykov (Moscow) 
and M. Nagano (Tokyo) for making the results of their 
respective experiments available in numerical form. 
I am also grateful to P. Grieder (Bern) and L. Jones 
(Ann Arbor) for their help in retrieving literature 
and D. Heck and M. Roth (both at Karlsruhe) for their 
advice concerning LATEX and PAW (and their patience). 
Last not least I thank the Forschungszentrum Karlsruhe, 
especially H. Bl\"umer, for permission to use its 
facilities even after retirement.\\

\end{document}